\documentclass[twocolumn, times, tighten,twocolappendix]{aastex63}

\usepackage{graphicx,multirow,hyperref,url,color,xspace}

\newcommand{\msun}{{\rm M}_{\sun}}

\newcommand{\nustar}{{\textit{NuSTAR}}\xspace}
\newcommand{\nicer}{{\textit{NICER}}\xspace}
\newcommand{\integral}{{\textit{INTEGRAL}}\xspace}
\newcommand{\source}{{MAXI J1820+070}\xspace}

\defcitealias{Buisson19}{B19}
\newcommand{\buisson}{{\citetalias{Buisson19}}\xspace}
\defcitealias{Kara19}{K19}
\newcommand{\kara}{{\citetalias{Kara19}}\xspace}

\begin{document}

\title{Accretion Geometry in the Hard State of the Black Hole X-Ray Binary MAXI J1820+070}
\shorttitle{The hard state of MAXI J1820+070}

\author{Andrzej A. Zdziarski}
\affil{Nicolaus Copernicus Astronomical Center, Polish Academy of Sciences, Bartycka 18, PL-00-716 Warszawa, Poland; \href{mailto:aaz@camk.edu.pl}{aaz@camk.edu.pl}}
\author{Marta A. Dzie{\l}ak}
\affil{Nicolaus Copernicus Astronomical Center, Polish Academy of Sciences, Bartycka 18, PL-00-716 Warszawa, Poland; \href{mailto:aaz@camk.edu.pl}{aaz@camk.edu.pl}}
\author{Barbara De Marco}
\affil{Departament de F{\'{\i}}sica, EEBE, Universitat Polit{\`e}cnica de Catalunya, Av.\ Eduard Maristany 16, E-08019 Barcelona, Spain; \href{mailto:barbara.de.marco@upc.edu}{barbara.de.marco@upc.edu}}
\author{Micha{\l} Szanecki}
\affil{Faculty of Physics and Applied Informatics, {\L}{\'o}d{\'z} University, Pomorska 149/153, PL-90-236 {\L}{\'o}d{\'z}, Poland; \href{mailto:andrzej.niedzwiecki@uni.lodz.pl}{andrzej.niedzwiecki@uni.lodz.pl}}
\author{Andrzej Nied{\'z}wiecki}
\affil{Faculty of Physics and Applied Informatics, {\L}{\'o}d{\'z} University, Pomorska 149/153, PL-90-236 {\L}{\'o}d{\'z}, Poland; \href{mailto:andrzej.niedzwiecki@uni.lodz.pl}{andrzej.niedzwiecki@uni.lodz.pl}}

\shortauthors{Zdziarski et al.}

\begin{abstract}
We study X-ray spectra from the outburst rise of the accreting black-hole binary MAXI J1820+070. We find that models having the disk inclinations within those of either the binary or the jet imply significant changes of the accretion disk inner radius during the luminous part of the hard spectral state, with that radius changing from $>$100 to $\sim$10 gravitational radii. The main trend is a decrease with the decreasing spectral hardness. Our analysis requires the accretion flow to be structured, with at least two components with different spectral slopes. The harder component dominates the bolometric luminosity and produces strong, narrow, X-ray reflection features. The softer component is responsible for the underlying broader reflection features. The data are compatible with the harder component having a large scale height, located downstream the disk truncation radius, and reflecting mostly from remote parts of the disk. The softer component forms a corona above the disk up to some transition radius. Our findings can explain the changes of the characteristic variability time scales, found in other works, as being driven by the changes of the disk characteristic radii.
\end{abstract}

\section{Introduction}
\label{intro}

The standard model of accretion onto a black hole (BH) postulates the presence of a geometrically thin and optically thick disk \citep{SS73,NT73}. The disk is close to thermodynamic equilibrium and its emission can be approximated by a sum of local blackbodies with a color correction \citep{Davis05}. In BH X-ray binaries (XRBs), this emission peaks in $EF_E$ at $E\sim 1$\,keV. On the other hand, BH XRBs in their hard spectral state have the peak of their $E F_E$ emission at $E\sim 10^2$\,keV (e.g., \citealt{DGK07}), which cannot be explained by that model. Observations of such spectra prompted development of models of hot accretion disks, where the electron temperature is $kT_{\rm e}\sim 10^2$\,keV \citep{SLE76, NY94, Abramowicz95, YN14}. Those hot disks are postulated to exist below some radius, $R_{\rm in}$, and be surrounded by standard accretion disks. 

Then, alternative models explaining the hard X-ray spectra were developed. The thin disk can be covered by a hot corona \citep{GRV79, SZ94}. However, due to the cooling by the underlying disk \citep{HM91, PVZ18}, such coronae emit spectra with the photon index of $\Gamma\gtrsim 2$ (defined by $F_E\propto E^{1-\Gamma}$), which are too soft to explain the hard state, where $\Gamma<2$ is observed. This can be resolved if the corona is outflowing \citep{Beloborodov99}. Another model postulates instead the presence of a static point-like source on the rotation axis of the BH, the so-called lamppost \citep{Martocchia96}. The lamppost, if it exists, should be in some way connected to the jet, present in the hard state, though theoretical explanations of it appear insufficient so far \citep{Yuan19a,Yuan19b}. The presence of jets in the hard state implies the presence of poloidal magnetic fields \citep{BZ77, BP82, Liska20}, which can strongly modify the accretion solutions (e.g., \citealt{BK74, Narayan03, McKinney12, Cao13, Salvesen16}). 

\begin{table*}\centering
\caption{Observations of \source with \nicer and \nustar in the hard state during the outburst rise. 
}
\vskip -0.4cm                               
\begin{tabular}{cccccccccc}
\hline
Epoch & \nicer Obs. ID & Start time & Exposure  & \nustar Obs. ID & Start time & Exposure A & Exposure B\\
&& End time &[s]&& End time& [s]& [s] \\
\hline
1 & 1200120103 &2018-03-13T23:57:07&10691  \\
&&2018-03-14T23:25:12&& 90401309002 &2018-03-14T22:30:12& 11769 & 11981\\
  & 1200120104 &2018-03-15T00:39:50&6652&&2018-03-15T10:27:37& \\
&&2018-03-15T21:02:34\\
2 & 1200120106 &2018-03-21T09:21:01&4302&90401309006 &2018-03-21T07:18:35& 4540 & 4540 \\
&&2018-03-21T23:16:08&&&2018-03-21T16:10:13\\
3 & 1200120110 &2018-03-24T23:40:27&19083&90401309010 &2018-03-24T20:41:22& 2660 & 2801 \\
&&2018-03-25T23:14:46&&&2018-03-25T00:49:32\\
4 & 1200120130 &2018-04-16T01:54:03&6015&90401309013 &2018-04-16T22:51:45& 1834 & 1934 \\
&&2018-04-16T23:50:30&&&2018-04-17T01:23:09\\
\hline
\end{tabular}
\label{log}
\end{table*}

Deciding which model actually applies to the hard state requires strong observational constraints. In particular, an accurate determination of the geometry of the inner accretion flow is crucial. So far, such determinations have given conflicting results. A large number of papers claim the disk in the hard state extends to the immediate vicinity of the innermost stable circular orbit (ISCO), while some other papers find the disk to be truncated (\citealt{Bambi20} and references therein). Here, we consider X-ray observations of the recent outburst of \source, a bright transient BH XRB, with the goal of resolving this controversy. We use data from two very sensitive instruments, {\it Nuclear Spectroscopic Telescope Array} (\nustar; \citealt{Harrison13}), and {\it Neutron star Interior Composition ExploreR} (\nicer; \citealt{Gendreau16}). We re-examine the findings by \citet{Kara19}, hereafter \kara, and \citet{Buisson19}, hereafter \buisson, that the source evolution in the hard state is dominated by a corona vertical contraction accompanied by an approximate constancy of the surrounding disk.

\source was discovered in 2018, first in the optical range \citep{Tucker18}, and five days later in X-rays \citep{Kawamuro18}. It is a relatively nearby source; its most accurate distance estimate appears to be the radio-parallax determination \citep{Atri20}, $d\approx 3.0\pm 0.3$\,kpc.  This is consistent with determinations based on the Gaia Data Release 2 parallax, $d\approx 3.5^{+2.2}_{-1.0}$\,kpc \citep{Bailer18,Gandhi19, Atri20}. The inclination of the binary has been estimated as $66\degr<i_{\rm b}<81\degr$ \citep{Torres19,Torres20}, while that of the jet, as $i_{\rm j}\approx 63\pm 3\degr$ \citep{Atri20}. The BH mass is anticorrelated with $i_{\rm b}$, $M\approx (5.95\pm 0.22)\msun/\sin^3 i_{\rm b}$ \citep{Torres20}. The high inclination of this source is confirmed by the detection of X-ray dips \citep{Kajava19}. 

\section{Observations and data reduction}
\label{data}

\begin{figure}[t!]
  \centering  \includegraphics[width=7.cm]{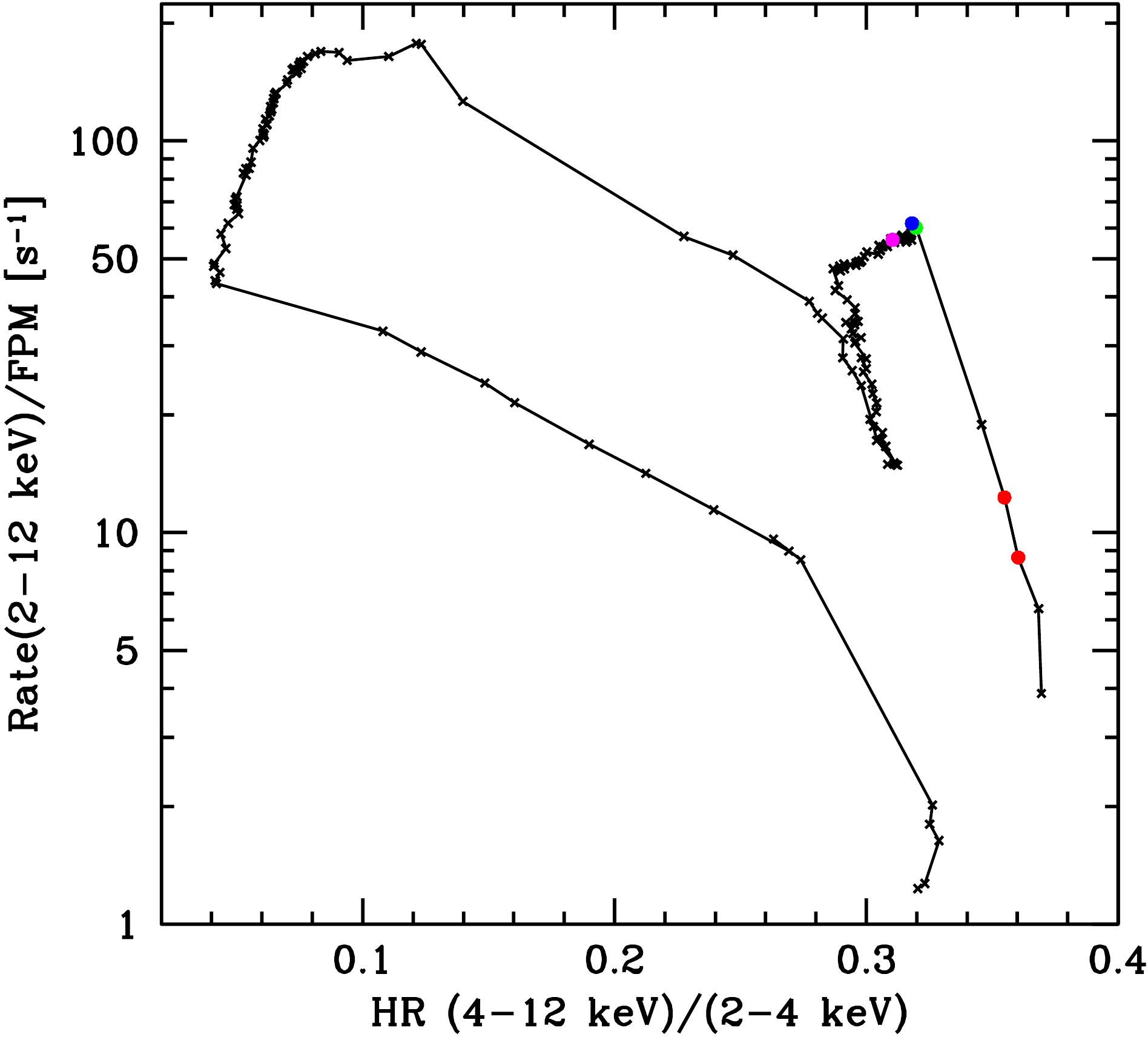}
  \caption{The \nicer count rate per a Focal Plane Module in the 2--12\,keV range vs.\ the hardness given by the count-rate ratio of 4--12 to 2--4\,keV for the main part of the outburst, between 2018-03-12T13:51:20 (MJD 58189.577; the rightmost point) and 2018-10-13T02:37:28 (MJD 58404.109; the lowest point). The lines connect observations adjacent in time. The \nustar observations during our epochs 1, 2, 3, 4 are contemporaneous to the \nicer observations indicated by the red, green, blue and magenta circles, respectively. 
}
\label{HCR}
\end{figure}

We have chosen the spectra from the hard state during initial phases of the outburst that have contemporaneous \nicer (0.3--12\,keV) and \nustar (3--79\,keV) coverage. The selected spectral data for four epochs are detailed in Table \ref{log}. However, we present here only the results of spectral fits to the \nustar data. The reason for this is that \nicer is primarily an instrument for timing studies, and its spectral calibration remains much less accurate than that of \nustar. In fact, fitting \nicer data requires introducing artificial edges to account for sharp instrumental residuals (e.g., \citealt{Wang20}), which significantly affects the accuracy of the fits. In this Letter, we use the joint data to measure the bolometric flux and to show the form of the soft excess below 3 keV. We intend to perform spectral fits of the joint data with an improved \nicer spectral calibration in a forthcoming paper.

We also show the count-rate vs.\ hardness diagram for 2018 outburst using the \nicer data in Figure \ref{HCR}. We see several phases. The initial rise in the hard state contains our epoch 1, as shown by the red circles. The rise reaches a local maximum, at which our epochs 2 and 3 are located. This was followed by a plateau, during which the count rate only slightly decreased during a decrease of the hardness, and our epoch 4 is located there. Subsequently, there was a rate decline associated with a hardening and a return along a similar path. This was followed by a transition to the soft state, after which the source returned to the hard state but at much lower fluxes. The four \nustar spectra considered by us were studied by \buisson, where they are denoted as epoch 1, the second parts of epochs 2 and 3, and the first part of epoch 5. Also, the \nustar data of our epoch 3 were studied in \citet{Chakraborty20}. Timing properties of the \nicer observations of our epochs 2 and 4 were studied in \kara. The \nicer data of epoch 1 are studied in \citet{Dzielak21}. The data from two observations, one on the day preceding and one on the day following that \nustar observation, are added. A comprehensive study of the timing properties of all of the \nicer observations from the phases of the outburst up to the transition to the soft state is given in \citet{DeMarco21}.

The \nustar data were reduced with {\sc heasoft} v.6.25, the {\tt NUSTARDAS} pipeline v.1.8.0, and {\tt CALDB} v.20200912. To filter passages through the South Atlantic Anomaly, we set {\tt saamode=strict} and {\tt tentacle=yes}. As recommended by the \nustar team, we use {\tt STATUS== b0000xxx00xxxx000} to avoid source photons being spuriously flagged as {\tt test}. The source region is a $60''$ circle centered on the peak brightness. The background is extracted from a $60''$ circle in an area with the lowest apparent contribution from sources. However, the background is negligible. We group the data to signal-to-noise (S/N) ratio $\geq$50, but to less at $>$69\,keV so to utilize the full $\leq$79\,keV band. 
 
The \nicer data were reduced using the {\tt NICERDAS} tools in {\sc heasoft} v.6.28 and {\tt CALDB} v.20200727. We applied the standard screening criteria \citep{Stevens18} and checked for periods of high particle background ($>$2\,s$^{-1}$) by using the 13--15\,keV light curves, where the source contribution is negligible \citep{Ludlam18}. We removed the Focal Plane Modules 14 and 34, which occasionally display increased noise, and screened for ones showing anomalous behavior. 

\section{Fits to the X-ray Spectra of \source}
\label{fits}

We study the spectra with the X-ray fitting package {\sc{xspec}} \citep{Arnaud96}. The reported fit uncertainties are for 90\% confidence, $\Delta\chi^2 \approx 2.71$. Residual differences between the calibration of the \nustar FPMA and B detectors are accounted for by the model {\tt jscrab} \citep{Steiner10}, which multiplies the spectrum by a power law with an index difference, $\Delta\Gamma$ (defined by the differential photon number flux of $\propto E^{-\Gamma}$) and a normalization. We account for the interstellar medium (ISM) absorption using the {\tt tbabs} model \citep{WAMC00} using the elemental abundances of \citet{AG89}. 

We use models with thermal Comptonization spectra incident on an accretion disk, taking into account atomic processes and relativistic effects, as implemented in two families of spectral codes, {\tt relxillCp, xillverCp} (v.\ 1.4.0; \citealt{GK10,Dauser16}) and {\tt reflkerr} \citep{Niedzwiecki19}. We assume a rotating BH with the dimensionless spin of $a_*=0.998$, for which $R_{\rm{ISCO}}\approx 1.237 R_{\rm{g}}$, where $R_{\rm g}\equiv GM/c^2$. At $R\gg R_{\rm{ISCO}}$, the metric is virtually independent of $a_*$. 

We begin with studying epoch 1. We first fit the \nustar spectra following \buisson, whose important conclusion was that no model with a single primary (Comptonization) component can fit the data. Their best model consists of a lamppost with two parts with the same incident spectra but at different heights, disk reflection normalized at that geometry, and a disk blackbody. The small differences of our analysis with respect to that work is that we use {\tt jscrab} instead of allowing independent disk blackbody parameters for the two FPMA and B. With the current calibration, $\Delta\Gamma$ of the FPMB with respect to FMBA is very small, $\approx\! +0.01$, and their differences at the lowest energies are similar to those typical for the entire range. Also, we include the ISM absorption, with $N_{\rm H}$ kept constant at $1.4\times 10^{21}$\,cm$^{-2}$ \citep{Kajava19,Dzielak21}. Thus, our model is {\tt jscrab*tbabs (diskbb+relxilllpCp$_1$+relxilllpCp$_2$)}, where the two {\tt relxilllpCp} terms give the two parts of the lamppost primary source. Finally, \buisson ignored the 11--12 and 23--28\,keV energy ranges, where there were some sharp instrumental features. We include those ranges since such features are no longer present.

\begin{figure}
  \centering  \includegraphics[width=7cm]{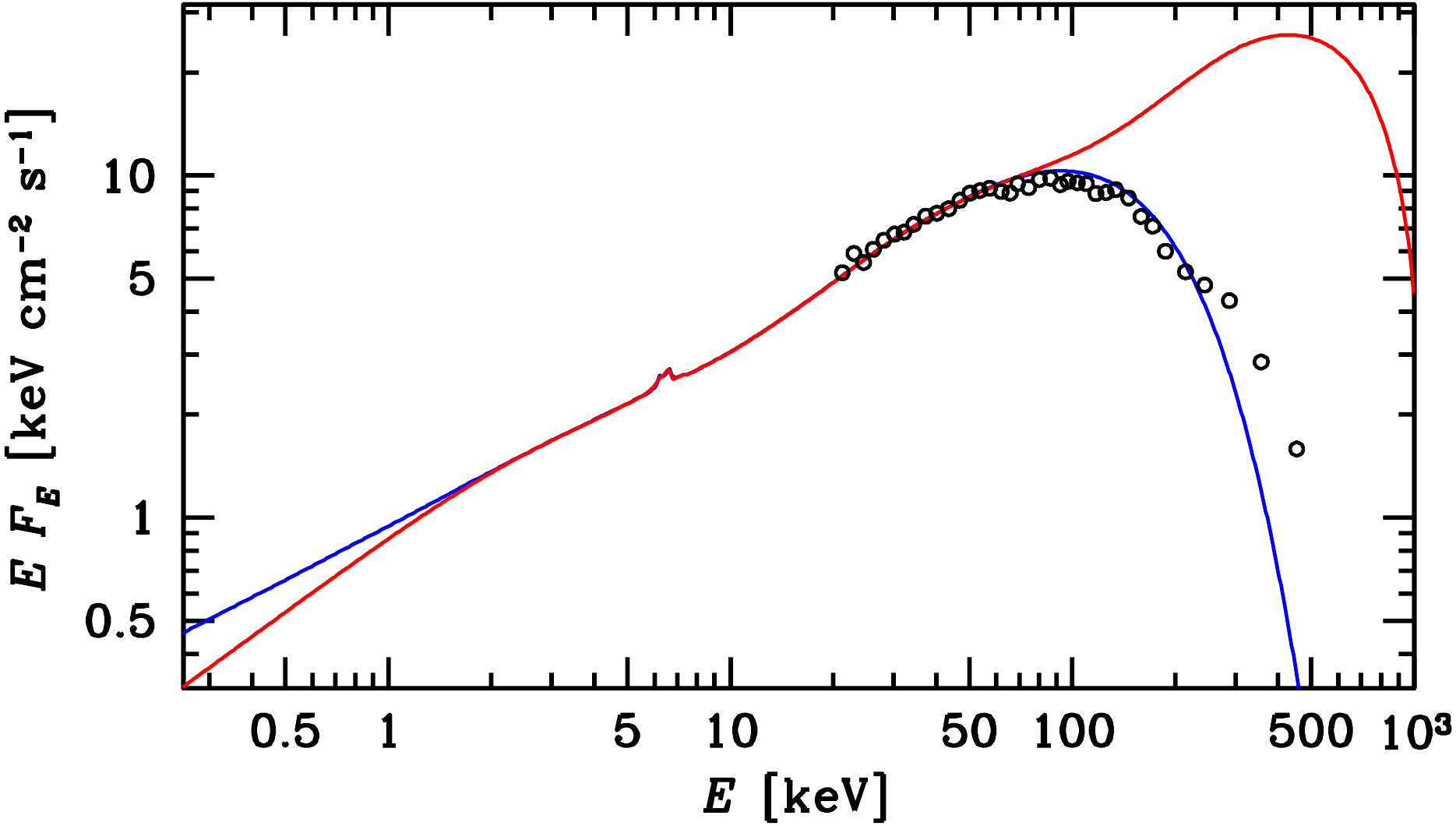}
  \caption{Comparison of the unabsorbed model spectra for the two fits of the double-lamppost + disk blackbody model to the epoch 1 data. The red and blue curves show the low and high $R_{\rm in}$ solutions, respectively. The two curves are virtually indistinguishable in the fitted 3--78\,keV range. The black circles show the measurements taken by the SPI detector on board \integral a day after the end of the \nustar observation \citep{Roques19}, normalized to the \nustar spectrum. We see that while it approximately agrees with the high-$R_{\rm in}$ model spectrum at $E>78$\,keV, it strongly disagrees with the low-$R_{\rm in}$ one.
}
\label{model1}
\end{figure}

\begin{figure}[t!]
  \centering  \includegraphics[width=7cm]{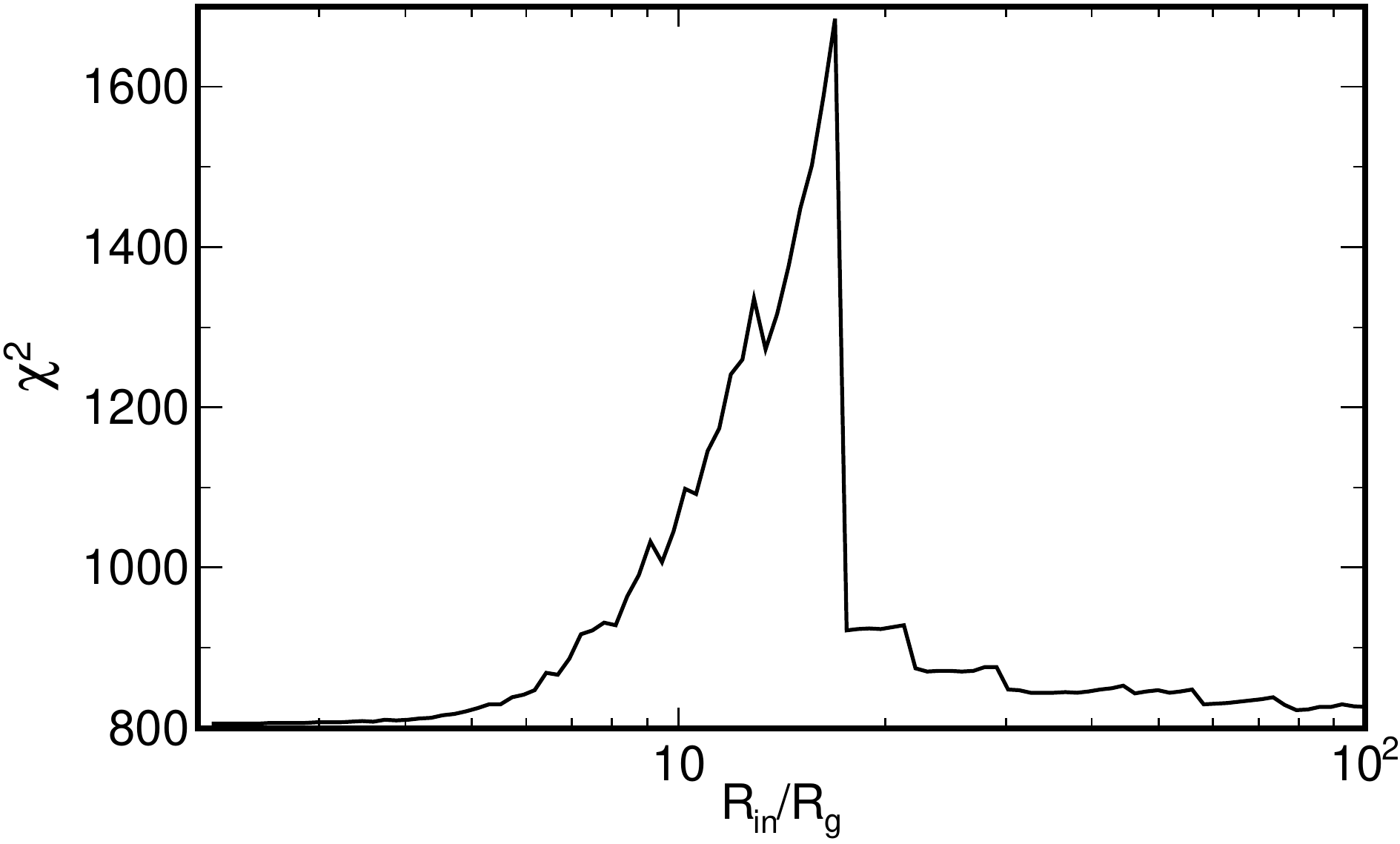}
  \caption{Dependence of $\chi^2$ on the disk inner radius, $R_{\rm in}$. We see two minima, at low and high values of $R_{\rm in}$, separated by a very high barrier in $\chi^2$, with $\Delta \chi^2\approx +900$ around $R_{\rm in}\approx 17 R_{\rm g}$.}
\label{chi2}
\end{figure}

We confirm the result of \buisson, with some small differences attributable to the updated \nustar calibration and a newer version of {\tt relxilllpCp}. We find a good fit with an extremely relativistic configuration, at $R_{\rm in}\approx 2.1^{+1.3}_{-0.5} R_{\rm g}$, $H_1 \approx 3.0^{+0.9}_{-0.1}R_{\rm g}$, $H_2 \approx 70^{+40}_{-30} R_{\rm g}$, a very hard spectral index, $\Gamma\approx 1.32^{+0.02}_{-0.01}$, at $\chi^2_\nu\approx 806/744$. Similar to \buisson, who found $i\approx 30^{+4}_{-5}\degr$, we obtain $i\approx 32^{+3}_{-5}\degr$. Given the current observational evidence for a high inclination (Section \ref{intro}), this represents a major problem for the applicability of this model (developed before the constraints on the inclination were published) to \source. Another problem for this spectral solution is that the Fe abundance is very high, $Z_{\rm Fe}\approx 6.1_{-0.2}^{+0.6}$ ($4.0^{+0.9}_{-0.7}$ in \buisson, which is unlikely given the presence of a weakly-evolved low-mass donor \citep{Torres20}. 

We have thus searched for alternative solutions at higher inclinations. We have indeed found a second minimum at a high inclination, though at somewhat higher $\chi^2_\nu\approx 820/744$. The inclination now fully consistent with the observational constraints, $i\approx 69^{+1}_{-9}\degr$. On the other hand, the disk is highly truncated, $R_{\rm in}\approx 77^{+200}_{-39} R_{\rm g}$. The height of the lower lamppost is $H_1 \approx 7.2^{+1.2}_{-5.2}R_{\rm g}$, that of the upper one is very large, $H_2\approx 500 R_{\rm g}$ (which is the largest value allowed by the adopted model), and $\Gamma\approx 1.50^{+0.01}_{-0.05}$. The observed emission is dominated by the upper lamppost. We have looked at possible degeneracies and significant correlations between the fitted parameters, but found none. 

The model spectra of the two solutions are compared in Figure \ref{model1}. We see that while the two curves are virtually indistinguishable in the fitted 3--78\,keV range, they make very different predictions at $\gtrsim$100\,keV.  The high-$R_{\rm in}$ fit predicts a gradual high-energy cutoff corresponding to the fitted electron temperature of $kT_{\rm e}\approx 58^{+3}_{-25}$\,keV. On the other hand, the low-$R_{\rm in}$ fit predicts a pronounced high-energy hump peaking at 0.4--0.5\,MeV, due to the high fitted $kT_{\rm e}\approx 360^{+40}_{-80}$\,keV ($400^{+0}_{-300}$\,keV in \buisson). The two spectra can be compared to the spectrum measured by the Spectrometer on \integral (SPI) about a day after the end of the \nustar observation, shown as the spectrum R1 in fig.\ 12 of \citet{Roques19}. We plot that spectrum in Figure \ref{model1}, multiplied by a factor of 0.52, which accounts for the flux increase during the time between the \nustar and SPI observations, and a calibration difference, with the SPI fluxes higher by a factor of $\approx$1.3 than those of \nustar for simultaneous observations. The error bars are not shown; they are of the order of the scatter among the points. We see that the SPI spectrum agrees well with that of \nustar in the overlapping 20--78\,keV range. It also approximately agrees at higher energies with the high-$R_{\rm in}$ model, but it disagrees with the low-$R_{\rm in}$ one. This provides one more argument against its physical reality. We have also compared the residuals of the two fits. We found them to be very similar, with no systematic differences (which is compatible with the close similarity of the shape of the best-fit models, shown in Figure \ref{model1}).

We have also found that the two solutions are separated by a very high barrier in $\chi^2$, as shown in Figure \ref{chi2}. The barrier reaches $\Delta \chi^2\approx +900$ around $R_{\rm in}\approx 17 R_{\rm g}$. This actually prevents finding the low-$R_{\rm in}$ solution when starting from the high-$R_{\rm in}$ one, using either {\tt steppar} in {\sc xspec}, or the Monte Carlo Markov Chain method. We have also looked for low-$R_{\rm in}$ counterparts of other spectral solutions presented below, and found they are generally present, but their inclinations are also at $i\sim 30\degr$. 

The analyses of the binary parameters \citep{Torres19,Torres20} and of the jet \citep{Atri20}, and the presence of X-ray dips (\citealt{Kajava19}; see a discussion in \citealt{FKR02}) all show at very high confidence that the source inclination is high. The inner disk can be aligned either with the binary plane or the normal to the BH rotation axis, but both are inclined at $\gtrsim\!60\degr$. An outer part of the disk could be warped (e.g., \citealt{Pringle97}), but not the disk in an immediate vicinity of the ISCO. Therefore, while the $i\sim 30\degr$ spectral solution is statistically better than the $i\sim 60$--$70\degr$ one, it can be considered only as a phenomenological description of the spectrum, but not as a representation of the actual geometry of the accretion flow. Therefore, in the remainder of this study, we consider only high-$i$ solutions. 

Furthermore, the disk blackbody component, included in this model in order to account for the soft excess present in the $\geq$3\, keV data, is also phenomenological, and its inner temperature, $kT_{\rm in}\approx 1.4^{+0.1}_{-0.2}$\,keV, is significantly higher than that seen in the \nicer data, which is $kT_{\rm in}\sim 0.2$\,keV \citep{Wang20_HXMT,Dzielak21}, which component contributes negligibly at energies $>$3\,keV. On the other hand, in the framework of such a composite lamppost, the two incident spectra should have different spectral indices. These spectra are most likely from the Comptonization process, which implies that the spectral slope at $E\ll kT_{\rm e}$ is a sensitive function of the flux of the incident seed soft photons from the surrounding accretion disk, and of the magnetic field strength if the cyclo-synchrotron process is important. Both strongly depend on the height of the X-ray source in the assumed geometry. Furthermore, the large fitted $R_{\rm in}$ suggests the possibility that the region downstream of it contains a hot plasma, and the adopted lamppost model is a proxy to a more complex physical situation. Therefore, we allow the reflection strengths to be free parameters.  

Thus, we consider a model with the two incident spectra having different parameters. In this case, the presence of a disk blackbody is not required. We have found a similar $\chi^2_\nu$, $\approx 820/743$, and $R_{\rm in}\approx 107^{+172}_{-95} R_{\rm g}$, $i\approx 61^{+9}_{-1}\degr$, $Z_{\rm Fe}\approx 2.0_{-0.5}^{+0.2}$. The lower lamppost component is soft, with $\Gamma_1\approx 1.80^{+0.06}_{-0.27}$, at a low height, $H_1 \approx 2.5^{+0.1}_{-0.4} R_{\rm g}$, and the part of the disk giving rise to most of the reflection is strongly ionized, $\log_{10}\xi\approx 4.3^{+0.2}_{-0.1}$ (where the ionization parameter, $\xi$ is in units of erg\,cm\,s$^{-1}$). The upper lamppost has $\Gamma_2\approx 1.44^{+0.01}_{-0.01}$, $H_2 \approx 500 R_{\rm g}$, and its reflecting part of the disk is weakly ionized, $\log_{10}\xi\approx 0.3^{+1.5}_{-0.3}$ (where the lower limit corresponds to the minimum allowed in the model). We find that the upper lamppost dominates both the bolometric flux and the observed, relatively narrow, Fe K complex. Given the flux dominance and the \nustar energy coverage limited to $<$78\,keV, we assume $kT_{\rm e}$ to be the same for both components, and find it $\approx 47_{-8}^{+8}$\,keV. 

These results bring about the issue of the physical nature of the spectral components. Likely, the region at $R<R_{\rm in}$ is filled by a hot plasma, which irradiates the truncated disk. However, the data clearly require two separate plasma clouds. The one radiating most of the luminosity also dominates the observed reflection features, which are almost non-relativistic. The presence of such features in the data implies the reflection from remote parts of the disk, which drives the large upper lamppost height, $\sim\! 500 R_{\rm g}$. Indeed, a similar fit can be obtained when replacing the upper lamppost by a static reflection component, {\tt xillverCp}.

\begin{figure}
  \centering  \includegraphics[width=7.5cm]{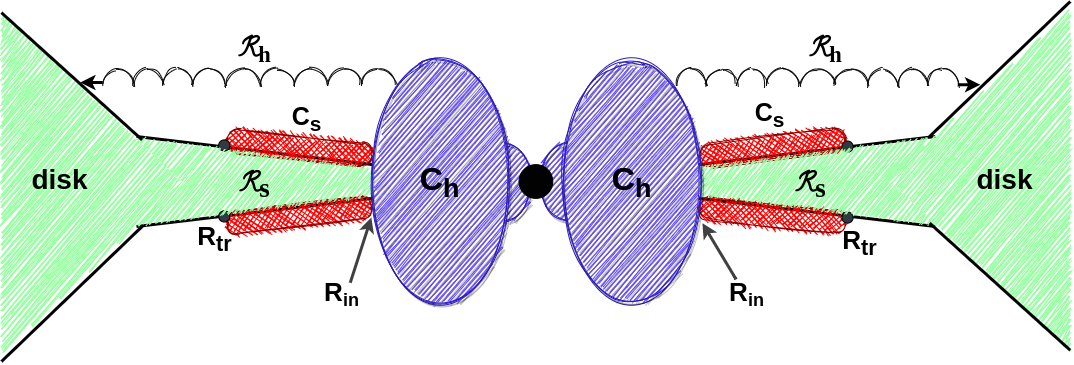}
  \caption{Schematic representation of the proposed geometry for the spectral fits shown below in Figure \ref{reflkerr2}. The disk is truncated at $R_{\rm in}$, and covered by a Comptonizing coronal plasma, $C_{\rm s}$, from $R_{\rm in}$ to $R_{\rm tr}$. Interior to $R_{\rm in}$, there is a hot accretion flow with a relatively large scale height, $C_{\rm h}$. Comptonization in $C_{\rm s}$ and $C_{\rm h}$ gives rise to the observed softer and harder, respectively, incident spectral components. The emission of $C_{\rm h}$ is reflected from the flared disk beyond $R_{\rm tr}$, marked as ${\cal R}_{\rm h}$. The emission of the coronal plasma, $C_{\rm s}$, is reflected from the disk beneath it, marked as ${\cal R}_{\rm s}$.
}
\label{geo}
\end{figure}

Following these results, we have developed an accretion model with two hot plasma flows and an accretion disk, shown in Figure \ref{geo}. The disk is truncated at $R_{\rm in}$, and significantly flared at large radii (as follows from the standard accretion models, e.g., \citealt{SS73}) and/or warped. The disk is covered by a hot Comptonizing corona, $C_{\rm s}$, from $R_{\rm in}$ to $R_{\rm tr}$, which is responsible for the observed softer incident spectral component. Its emission is reflected from the underlying disk, which is strongly ionized and which reflection is partly attenuated by the subsequent scattering in the plasma. (For the sake of simplicity, we assumed that this only results in a reduction of the observed reflection strength.) Interior to $R_{\rm in}$, there is also a hot accretion flow, $C_{\rm h}$, which Comptonization gives rise to the observed harder incident component. Since the disk is covered by the hot corona up to $R_{\rm tr}$, this emission is reflected predominantly by the bare disk beyond it. That region is much less ionized given the large distance between the plasma and the reflector. 

However, we note that the above scenario requires $R_{\rm tr}>R_{\rm in}$, i.e., it forces the inner radius of the hard reflection component to be larger than that for the soft one. In order to test the validity of this assumption, we also consider an alternative model allowing for an overlap of the reflection regions. This can correspond to the hot corona between $R_{\rm in}$ and $R_{\rm tr}$ being patchy. In this version, we fit two values of the inner radii of the reflecting region, $R_{\rm in,s}$ and $R_{\rm in,h}$, without imposing any a priori conditions on them. 

\begin{table*}
\caption{The results of spectral fitting for our two-component coronal model, {\tt jscrab*tbabs(reflkerr$_{\rm s}$+reflkerr$_{\rm h}$)}, to the \nustar data, and for two options, separate and overlapping reflection regions. 
}
  \centering\begin{tabular}{cccccc}
\hline
Component & Parameter & Epoch 1 & Epoch 2 & Epoch 3 & Epoch 4\\
\hline
ISM absorption & $N_{\rm H}$ $[10^{21}]$\,cm$^{-2}$ & \multicolumn{4}{c}{1.4f}\\
\hline
\multicolumn{6}{c}{\bf Separate reflection regions}\\
\hline
\hline
Joint constraints & $i$ $[\degr$] & $63^{+2}_{-2}$ & $66^{+1}_{-1}$ & $59_{-10}^{+7}$ & $71^{+1}_{-9}$\\
& $Z_{\rm Fe}$ & $1.3^{+0.1}_{-0.1}$ & $1.2^{+0.2}_{-0.1}$ & $1.4^{+0.2}_{-0.3}$ & $1.2^{+0.2}_{-0.3}$\\
& $kT_{\rm e}$ [keV] & $44_{-1}^{+3}$ & $27_{-2}^{+2}$ & $24_{-1}^{+1}$ & $32_{-4}^{+4}$\\
\hline
Soft Comptonization  & $y_{\rm s}$ & $0.98^{-0.05}_{+0.07}$ & $1.11_{-0.03}^{+0.02}$ & $1.17_{-0.02}^{+0.02}$& $0.81^{-0.09}_{+0.09}$\\
and reflection  &$R_{\rm in}\, [R_{\rm g}]$ & $77_{-24}^{+335}$ & $16_{-7}^{+6}$ & $127_{-61}^{+71}$ & $10.1_{-3.6}^{+4.7}$\\
& ${\cal R}_{\rm s}$ & $0.61_{+0.13}^{-0.32}$ & $0.57^{+0.18}_{-0.13}$ & $0.37^{+0.09}_{-0.17}$ & $0.86^{+0.27}_{-0.38}$\\
 & $\log_{10} \xi_{\rm s}$ & $3.74_{-0.53}^{+0.55}$ & $3.48_{-0.10}^{+0.10}$ & $3.49_{-0.12}^{+0.12}$ & $3.30_{-0.22}^{+0.09}$\\
& $N_{\rm s}$ & 0.18 & 2.57 & 3.47 & 1.70\\
& $F_{\rm s,inc}\,[10^{-8}{\rm erg/cm}^{2}\,{\rm s}]$ & 0.23 & 4.1 & 5.9 & 1.5\\
\hline 
Hard Comptonization &$y_{\rm h}$ & $1.78_{-0.02}^{+0.01}$ & $1.71_{-0.06}^{+0.04}$ & $1.89_{-0.09}^{+0.07}$ & $1.41_{-0.07}^{+0.04}$\\
and reflection &$\Delta R\, [R_{\rm g}]$ & $230_{-220}^{+120}$ & $170_{-50}^{+60}$ & $\leq$110 & $57_{-19}^{+35}$\\ 
& $R_{\rm tr}\, [R_{\rm g}]$ & \multicolumn{4}{c}{$=R_{\rm in}+\Delta R$}\\
&$R_{\rm out}\, [R_{\rm g}]$ & \multicolumn{4}{c}{$10^3$f}\\
& ${\cal R}_{\rm h}$ & $0.32^{+0.02}_{-0.01}$ & $0.62^{+0.21}_{-0.06}$ & $0.62^{+0.38}_{-0.15}$ & $0.48^{+0.38}_{-0.03}$\\
& $\log_{10} \xi_{\rm h}$ & $0.30_{-0.06}^{+0.02}$ & $0.53_{-0.53}^{+1.39}$ & $0.10_{-0.10}^{+1.60}$ &$0.46_{-0.46}^{+1.54}$\\
& $N_{\rm h}$ & 0.80 & 1.60 & 0.97 & 3.30\\
& $F_{\rm h,inc}\,[10^{-8}{\rm erg/cm}^{2}\,{\rm s}]$ & 3.8 & 6.0 & 4.1 & 9.2\\
\hline
& $\chi_\nu^2$  & 818/744 & 1040/879 & 704/699 & 703/543\\
\hline
\hline
\multicolumn{6}{c}{\bf Overlapping reflection regions}\\
\hline
\hline
Joint constraints & $i$ $[\degr$] & $63^{+2}_{-2}$ & $66^{+1}_{-1}$ & $63_{-3}^{+3}$ & $71^{+1}_{-9}$\\
& $Z_{\rm Fe}$ & $1.4^{+0.2}_{-0.1}$ & $1.1^{+0.2}_{-0.1}$ & $1.3^{+0.3}_{-0.2}$ & $1.2^{+0.3}_{-0.2}$\\
& $kT_{\rm e}$ [keV] & $44_{-6}^{+3}$ & $26_{-2}^{+1}$ & $24_{-1}^{+1}$ & $31_{-3}^{+4}$\\
\hline
Soft Comptonization & $y_{\rm s}$ & $0.98^{-0.10}_{+0.05}$ & $1.04_{-0.02}^{+0.03}$ & $1.13_{-0.02}^{+0.02}$& $0.81^{+0.09}_{-0.07}$\\
and reflection &$R_{\rm in,s}\, [R_{\rm g}]$ & $60_{-52}^{+870}$ & $14_{-5}^{+6}$ & $95_{-56}^{+105}$ & $9.6_{-2.5}^{+4.7}$\\
 &$R_{\rm out,s}\, [R_{\rm g}]$ & \multicolumn{4}{c}{$10^3$f}\\
 & ${\cal R}_{\rm s}$ & $0.56_{+0.19}^{-0.15}$ & $0.75^{+0.61}_{-0.09}$ & $0.44^{+0.15}_{-0.11}$ & $0.82^{+26}_{-0.30}$\\
 & $\log_{10} \xi_{\rm s}$ & $3.74_{-0.56}^{+0.30}$ & $3.46_{-0.07}^{+0.07}$ & $3.44_{-0.05}^{+0.26}$ & $3.28_{-0.19}^{+0.10}$\\
& $N_{\rm s}$ & 0.18 & 2.02 & 3.27 & 1.78\\
& $F_{\rm s,inc}\,[10^{-8}{\rm erg/cm}^{2}\,{\rm s}]$ & 0.23 & 2.8 & 5.2 & 1.6\\
\hline 
Hard Comptonization &$y_{\rm h}$ & $1.78_{-0.01}^{+0.03}$ & $1.65_{-0.14}^{+0.05}$ & $1.86_{-0.09}^{+0.06}$ & $1.43_{-0.08}^{+0.05}$\\
and reflection & $R_{\rm in,h}\, [R_{\rm g}]$ & $290_{-80}^{+110}$ & $170^{+100}_{-40}$ & $130^{+100}_{-40}$ & $66^{+23}_{-17}$\\
&$R_{\rm out,h}\, [R_{\rm g}]$ & \multicolumn{4}{c}{$10^3$f}\\
& ${\cal R}_{\rm h}$ & $0.32^{+0.01}_{-0.01}$ & $0.50^{+0.09}_{-0.05}$ & $0.56^{+1.22}_{-0.03}$ & $0.47^{+0.43}_{-0.06}$\\
& $\log_{10} \xi_{\rm h}$ & $0.30_{-0.03}^{+0.02}$ & $0.43_{-0.05}^{+1.46}$ & $0.10_{-0.09}^{+1.68}$ &$0.42_{-0.42}^{+1.42}$\\
& $N_{\rm h}$ & 0.80 & 2.12 & 1.19 & 3.20\\
& $F_{\rm h,inc}\,[10^{-8}{\rm erg/cm}^{2}\,{\rm s}]$ & 3.8 & 7.5 & 5.1 & 8.9\\
\hline
& $\chi_\nu^2$  & 818/744 & 1041/879 & 704/699 & 703/543\\
\hline
\hline
& $F_{\rm bol}\,[10^{-8}{\rm erg/cm}^{2}\,{\rm s}]$ & 5.0 & 15 & 15 & 14 \\
& $L/L_{\rm E}$ & 0.046 & 0.14 & 0.14 & 0.13\\
\hline
\end{tabular}
\tablecomments{
See Section \ref{fits} for details. $kT_{\rm e}$ and $kT_{\rm bb}=0.2$\,keV are assumed to be the same for both Comptonizing coronae, $N$ is the flux density at 1\,keV in the observer's frame, and `f' denotes a fixed parameter. $F_{\rm (s,h),inc}$ give the (unabsorbed) bolometric Comptonization fluxes of the respective component, $F_{\rm bol}$ is an estimate of the total bolometric flux based on both the \nicer and \nustar data (normalized to the \nustar FPMA), and $L/L_{\rm E}$ is the Eddington ratio for $d=3$\,kpc, $M=8\msun$ and $X=0.7$ [$L_{\rm E}=1.47(M/\msun)\times 10^{38}$\,erg\,s$^{-1}$].}
\label{t_fits}
\end{table*}

\begin{figure}
\centering
\includegraphics[height=6.cm,angle=-90]{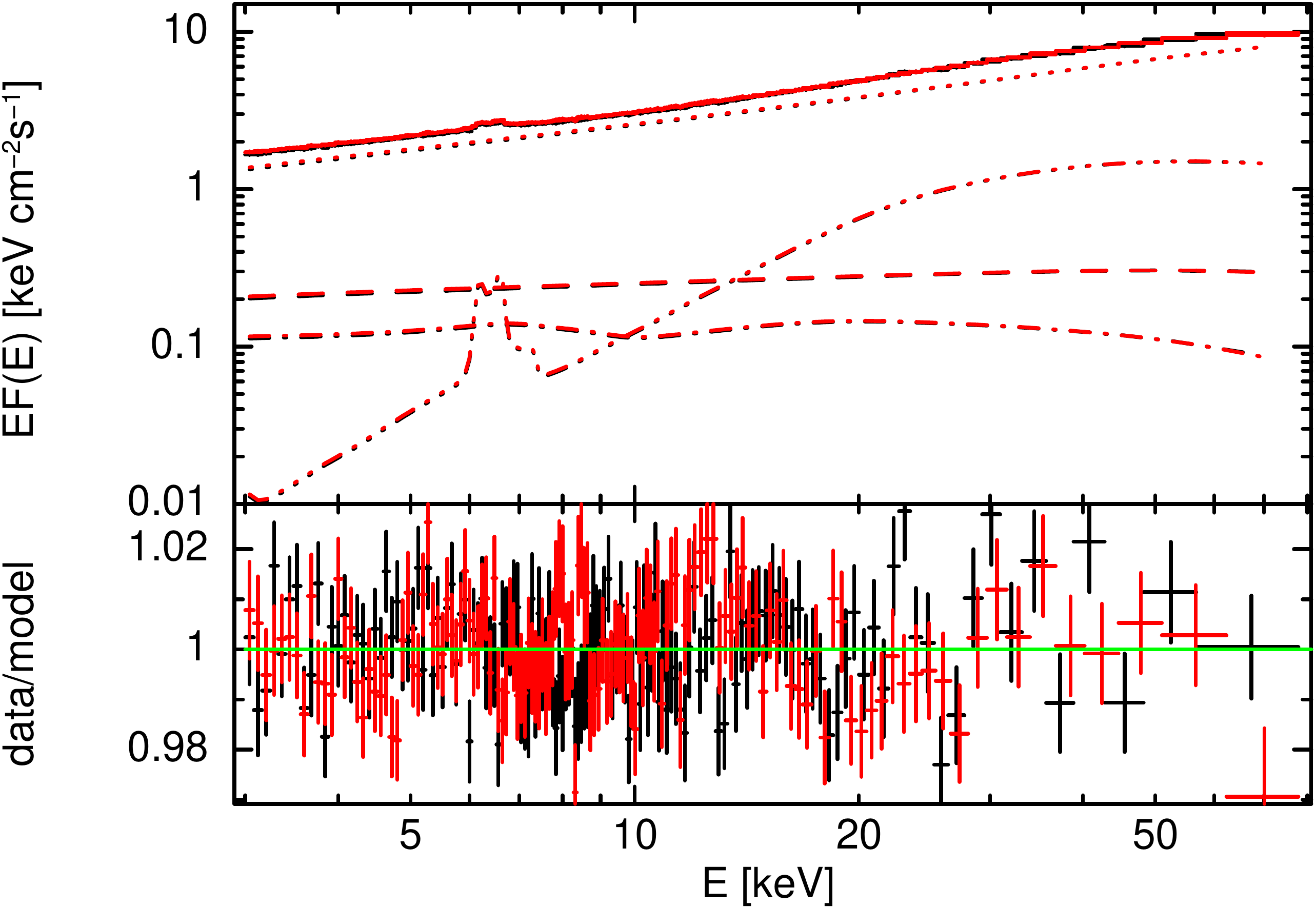} 
\includegraphics[height=6.cm,angle=-90]{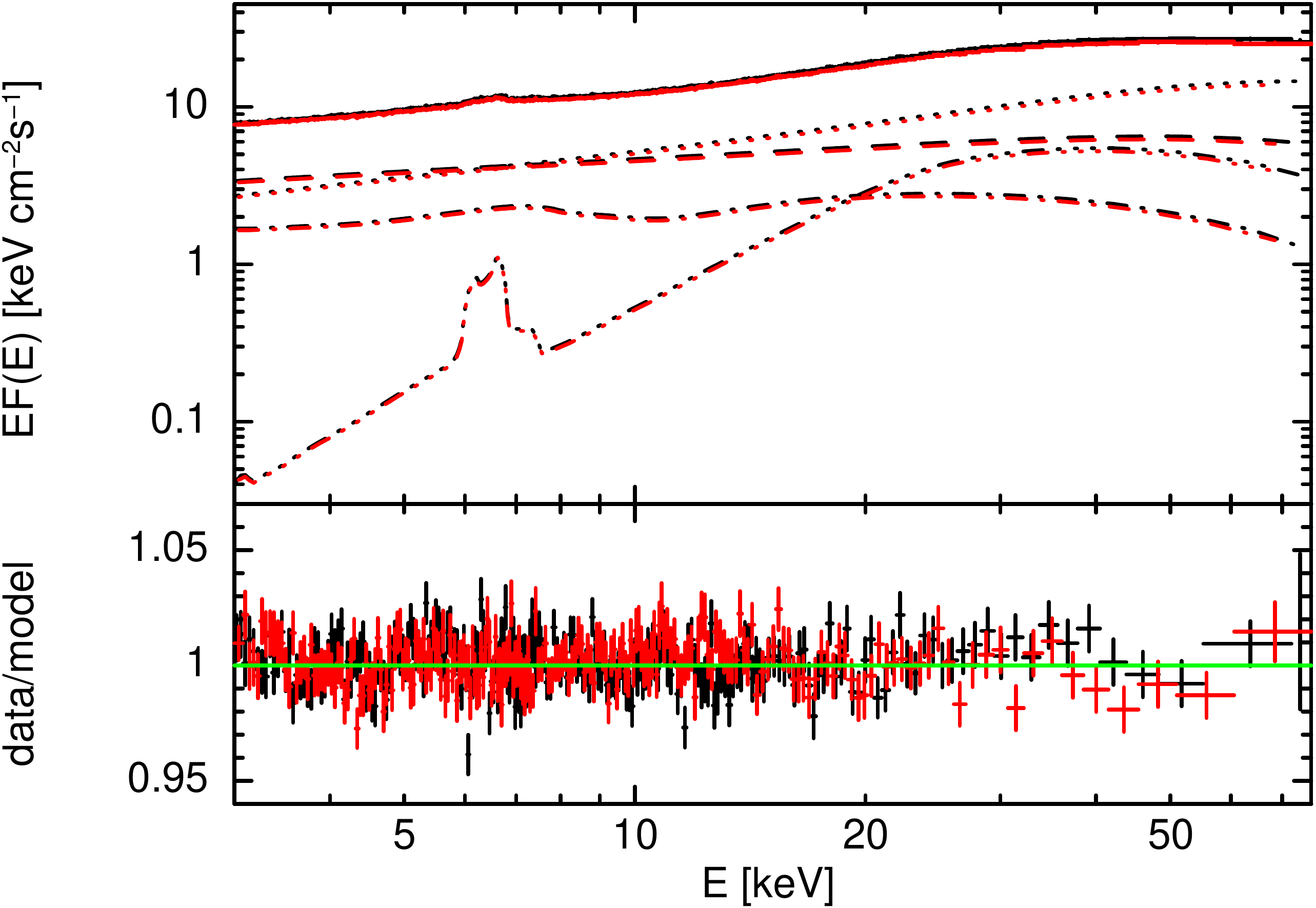}
\includegraphics[height=6.cm,angle=-90]{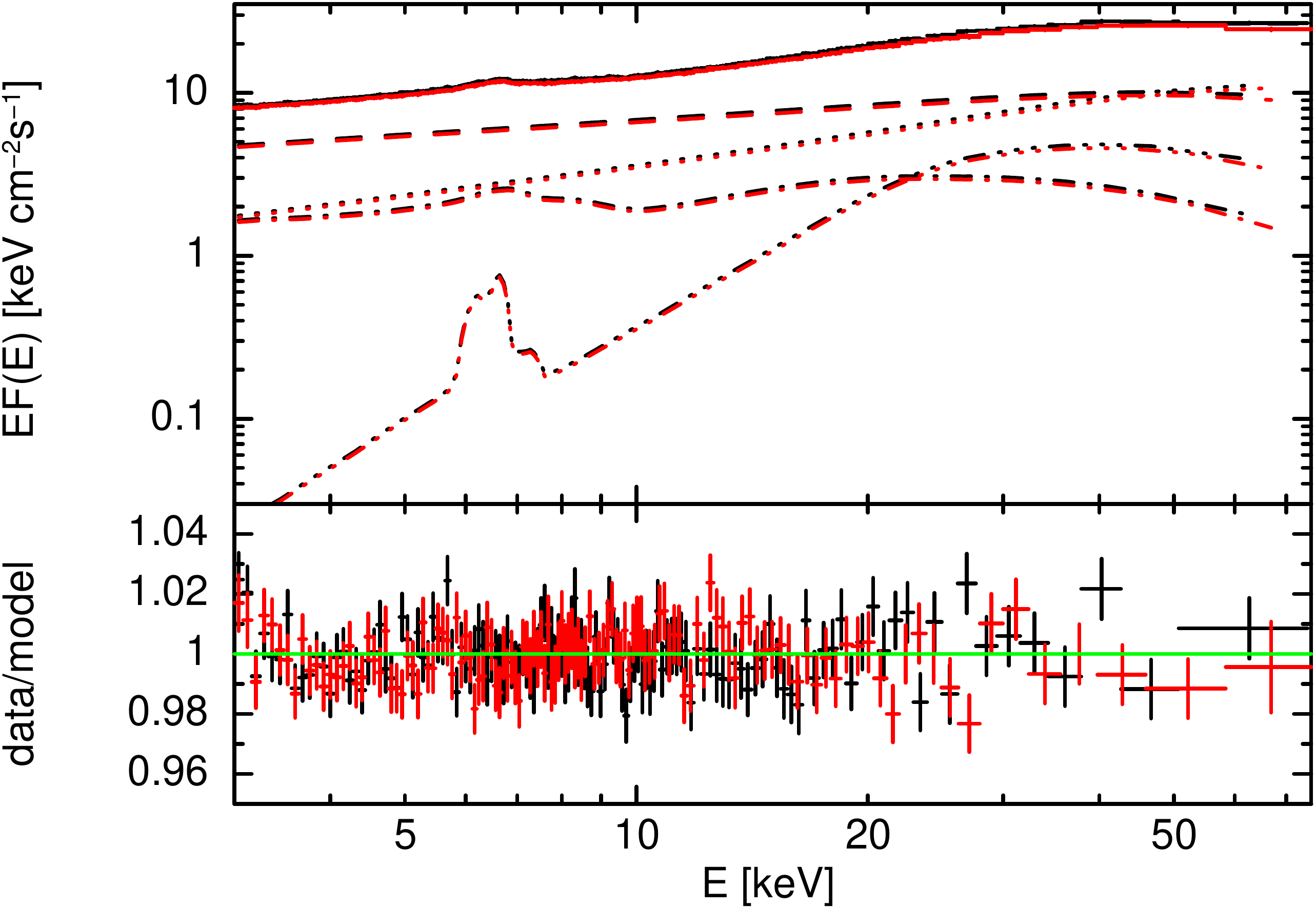} \includegraphics[height=6.cm,angle=-90]{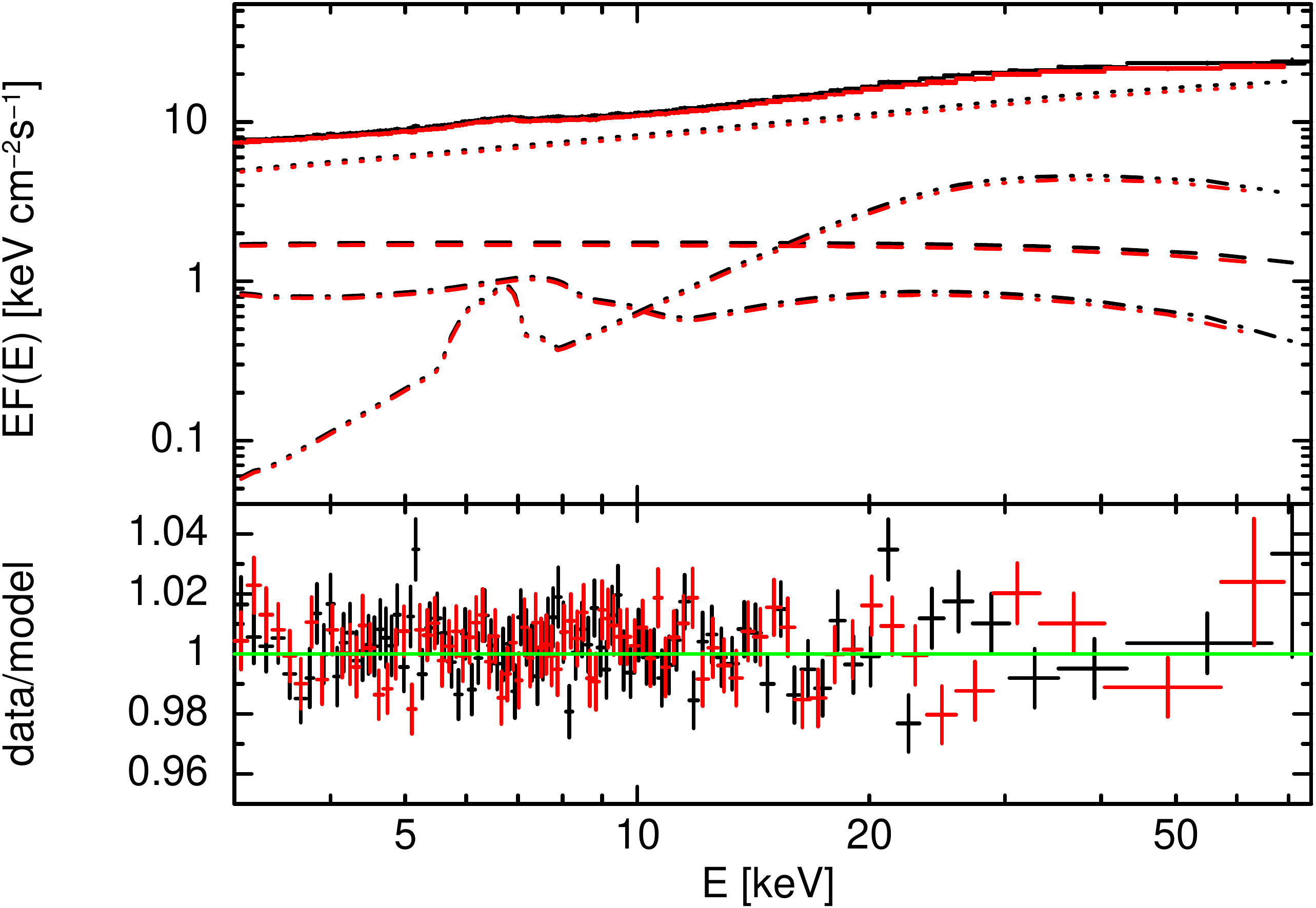}
  \caption{\nustar unfolded spectra and data-to-model ratios of the epochs 1--4 (from top to bottom) fitted in the 3--78\,keV range with the two-component coronal model (Table \ref{t_fits}). The dotted and dashed curves on the top panel correspond to the direct emission of the harder and softer corona, respectively, and the triple-dot-dashed and dot-dashed curves correspond to the reflection of harder and softer direct emission, respectively. Hereafter, the black and red symbols correspond to the FPMA and B, respectively, and the plotted spectra are rebinned to S/N $\geq$100.
}\label{reflkerr2}
\end{figure}

\begin{figure}
  \centering  
\includegraphics[width=6.cm]{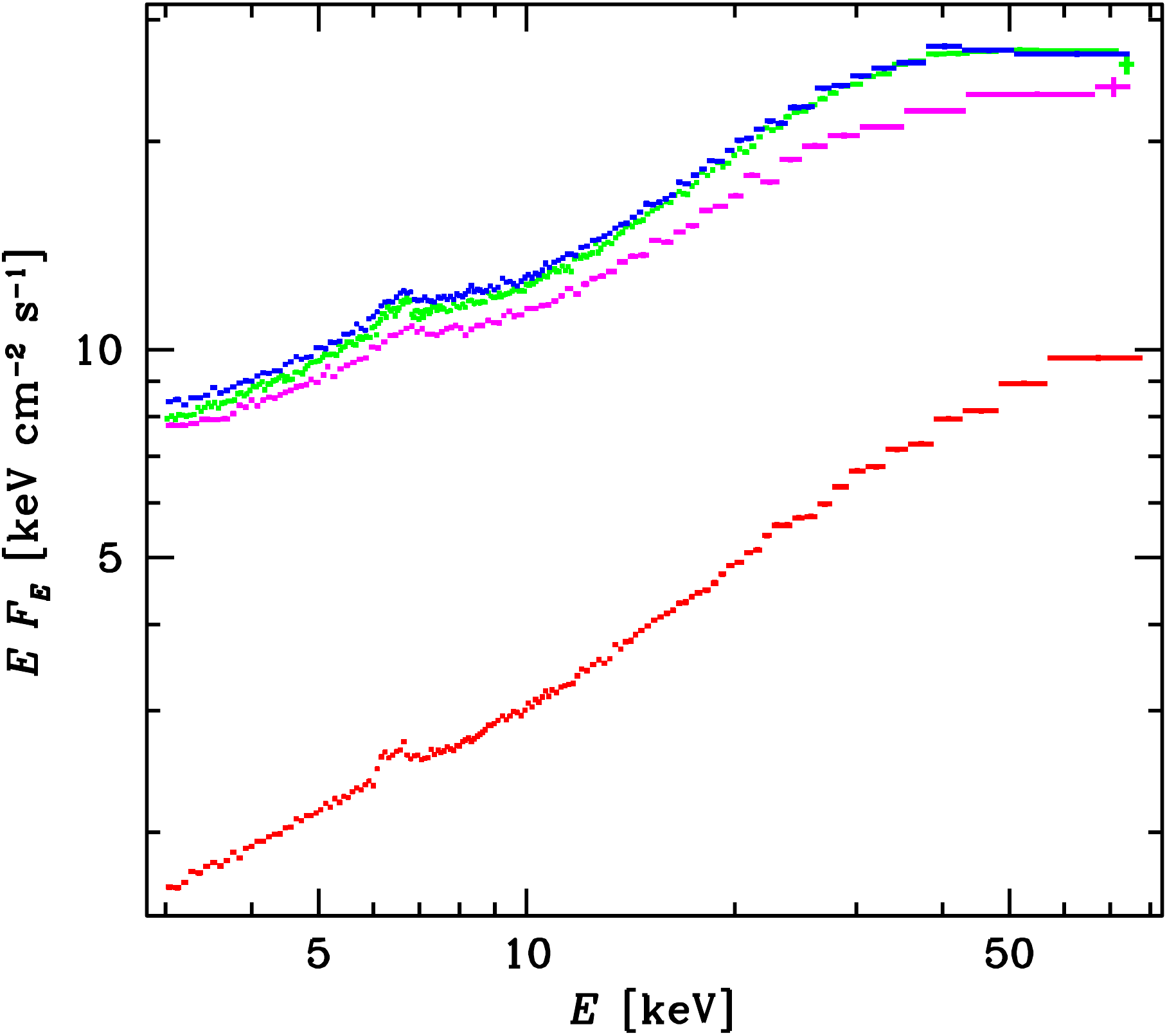}
  \caption{Spectra of the four studied observations unfolded with the two-component coronal model with separate reflection regions (Table \ref{t_fits}). For clarity, only the spectra from the FPMA detector are shown. The spectra of the epochs 1, 2, 3, 4 are shown in the red, green, blue and magenta color, respectively. 
}
\label{4spectra}
\end{figure}

\begin{figure}
\centering
\includegraphics[width=5.5cm]{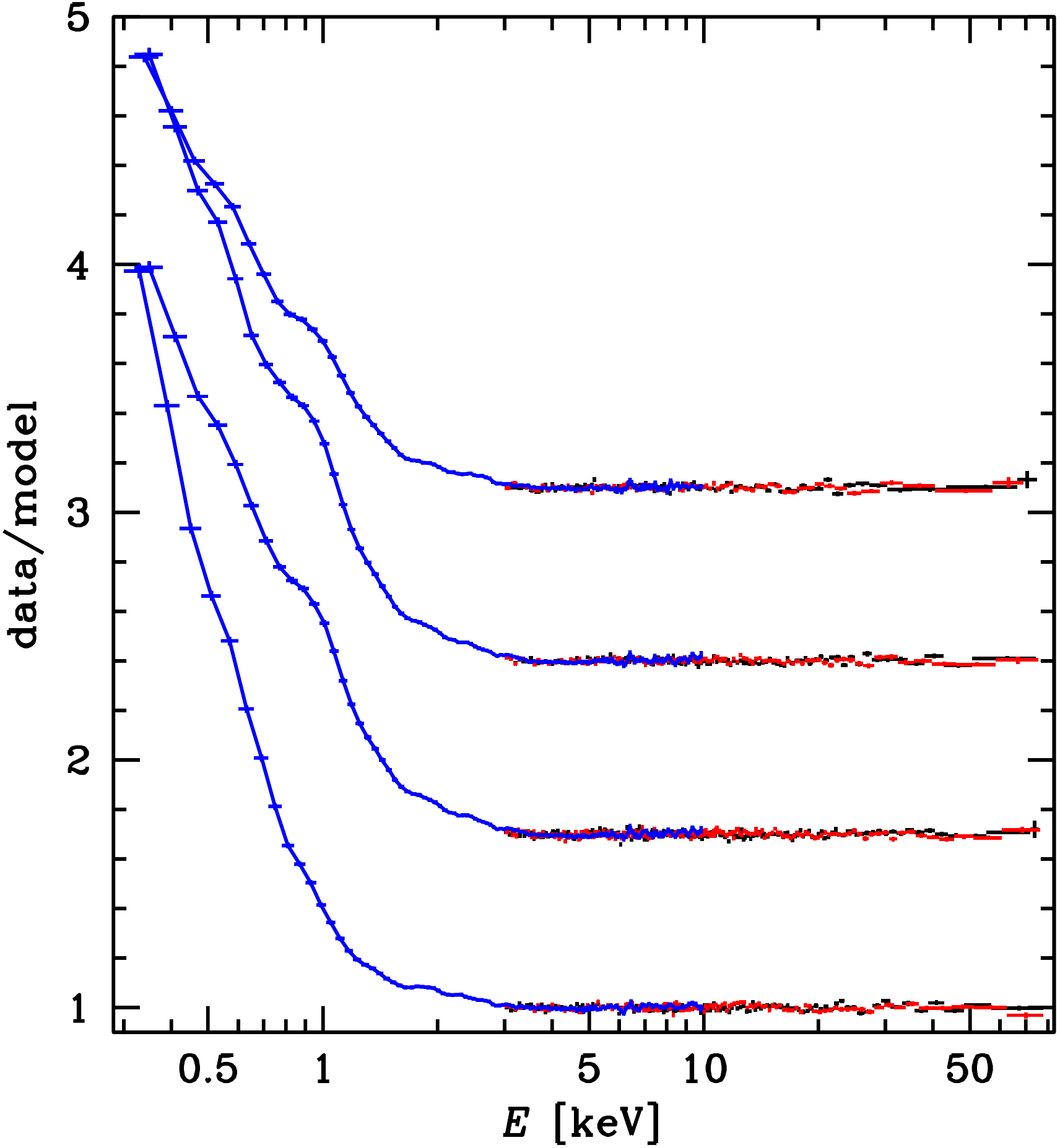} 
\caption{Data/model ratios for the two-component coronal model with separate reflection regions fitted to the \nustar (black and red symbols) and \nicer (blue symbols) 3--10\,keV data for epochs 1--4 (from bottom to top). For clarity of display, the profiles for epochs 2, 3 and 4 have been offset by $+0.7$, $+1.4$ and $+2.1$, respectively. We also show the \nicer data at $<$3\,keV, which show strong and complex soft excesses.}
\label{nustar_nicer_ratio}
\end{figure}

We implement the two variants of this geometry using the {\tt reflkerr} coronal reflection code of \citet{Niedzwiecki19}, which has a significantly improved treatment of Comptonization with respect to {\tt relxill}. As in {\tt relxill}, it approximates the reflection assuming a power-law disk irradiation profile; here we assume the standard profile of $\propto R^{-3}$, which follows the disk viscous dissipation at $R\gg R_{\rm ISCO}$. The used {\sc xspec} form is given in Table \ref{t_fits}. As previously, we assume $kT_{\rm e}$ to be the same for both components; allowing them to be different results only in a tiny reduction of $\chi^2$. The temperature of blackbody photons serving as seeds for Comptonization is taken as $kT_{\rm bb}=0.2$\,keV (compatible with the \nicer data; \citealt{Wang20_HXMT,Dzielak21}). The slope of each Comptonization component is parametrized by the Compton parameter, $y\equiv 4\tau_{\rm T} kT_{\rm e}/m_{\rm e}c^2$, where $\tau_{\rm T}$ is the Thomson optical depth of the plasma. We note, however, that the reported values of $y$ are somewhat overestimated due to the Comptonization model being based on iterative scattering \citep{PS96}. The reflection fraction, ${\cal R}$, is defined in {\tt reflkerr} as the ratio of the flux irradiating the disk to that emitted outside in a local frame. 

We find that both options of this model yield $\chi_\nu^2\approx 818/744$ and the parameters given in Table \ref{t_fits}. The two sets of the parameters differ only slightly. In the option with overlapping reflection regions, the inner radius of the soft component is lower than that for the hard component at the best fits for all four data sets, confirming the assumption of $R_{\rm tr}>R_{\rm in}$ done in the option with separate reflection regions (Figure \ref{geo}). In the latter, the value of $\Delta R\equiv R_{\rm tr}-R_{\rm in}$ is primarily driven by the hard reflection, as it determines $R_{\rm tr}$, while the soft reflection is relatively insensitive to its outer radius given the used irradiation profile $\propto R^{-3}$. The unfolded spectra and data/model ratios for the case of separate reflection regions are shown in Figure \ref{reflkerr2}. Both cases have the inclination within the observational constraints and the Fe abundance very close to solar. As expected if these fits indeed correspond to the geometry close to that shown in Figure \ref{geo}, the harder component dominates the bolometric flux, and the reflector of the softer component is much more ionized than that of the harder one. 

We have also tested a number of other models and found the truncation radius to be large and the reflection features are only weakly relativistic for all of those satisfying the observational constraints on the inclination. Thus, we have a robust conclusion that, at least in this BH XRB, the truncation radius at the luminosity of $\sim\! 5\% L_{\rm E}$ is $\sim\! 10^2 R_{\rm g}$.

We then study the data for epochs 2--4. We fit the \nustar spectra with the two versions of the double-corona {\tt reflkerr} model. We find good fits at high inclinations, as shown in Table \ref{t_fits} and Figure \ref{reflkerr2}. The shapes of the four spectra are compared in Figure \ref{4spectra}. The disk truncation radius is relatively small for epoch 2, $R_{\rm in}\sim 20 R_{\rm{g}}$ and even lower for 4, $R_{\rm in}\sim 10 R_{\rm{g}}$, while $R_{\rm in}$ during epoch 3 is similar to that of the epoch 1. In all cases, $R_{\rm in}$ is significantly larger than $R_{\rm ISCO}$. In epoch 3, we also find a high relative amplitude of the softer Comptonization component. We have no explanation for this feature of epoch 3; it may be due some fluctuation of the source parameters. We see a monotonous decrease of $R_{\rm tr}$ over all four epochs. Our values of $R_{\rm in}$ can be compared with those obtained by \citet{Wang20_HXMT} by modelling the \nicer data by a disk blackbody and a power law. They obtained the disk inner radii of $\approx (4.5$--$6.5)\times 10^7$\,cm during the period analyzed here. At $8\msun$, this corresponds to $\sim$40--$50 R_{\rm g}$, in an overall agreement with our values, and independently ruling out solutions with $R_{\rm in}$ close to the ISCO.

While the fits show variability of the characteristic disk radii, the fitted $i$ and $Z_{\rm Fe}$ are compatible with constant, as expected. In fact, we may expect some variability of $i$ due to precession or warping. Also, the viewing angle of the disk below and above $R_{\rm tr}$ may be different, as shown in Figure \ref{geo}. We neglect this complication as not to overfit the data. The values of our fitted inclination agree with the jet and binary observations, unlike those of \buisson and \citet{Chakraborty20}. The Fe abundance is compatible with $Z_{\rm Fe}\approx 1.2$--1.3. Such a closeness to unity is likely for this XRB, which has a low-mass donor \citep{Torres20}, in which substantial Fe synthesis is not expected during the evolution. On the other hand, the values of $Z_{\rm Fe}$ found by \buisson and \citet{Chakraborty20} were within the range of 4--10 at their best fits.

We then briefly consider the \nicer data. We fit them at the range 3--10\,keV, to test their consistency with the \nustar data. We find a generally good agreement, even in the relative normalization, which values are close to unity, and joint fits with the double-corona model give parameters similar to those in Table \ref{t_fits}. However, the \nicer data at $<$3\,keV show strong soft excesses, as shown in Figure \ref{nustar_nicer_ratio}. These excesses are not compatible with the presence of a disk blackbody alone, and imply a further complexity of the accretion flow (see \citealt{Dzielak21}). The joint data will be studied in a forthcoming paper, including the effect of quasi-thermal re-radiation of a fraction the incident flux \citep{ZDM20, Zdziarski21}. Now, we use them only to estimate the bolometric fluxes of the observations. We utilize a phenomenological Comptonization/reprocessing model to describe the overall shape of the broad band spectra and to estimate the fluxes below 3 keV, while we use our double-corona fits to get the fluxes above 3 keV. The resulting values are given in Table \ref{t_fits}.

\section{Discussion}
\label{discussion}

\begin{figure}
\centering
\includegraphics[width=6.5cm]{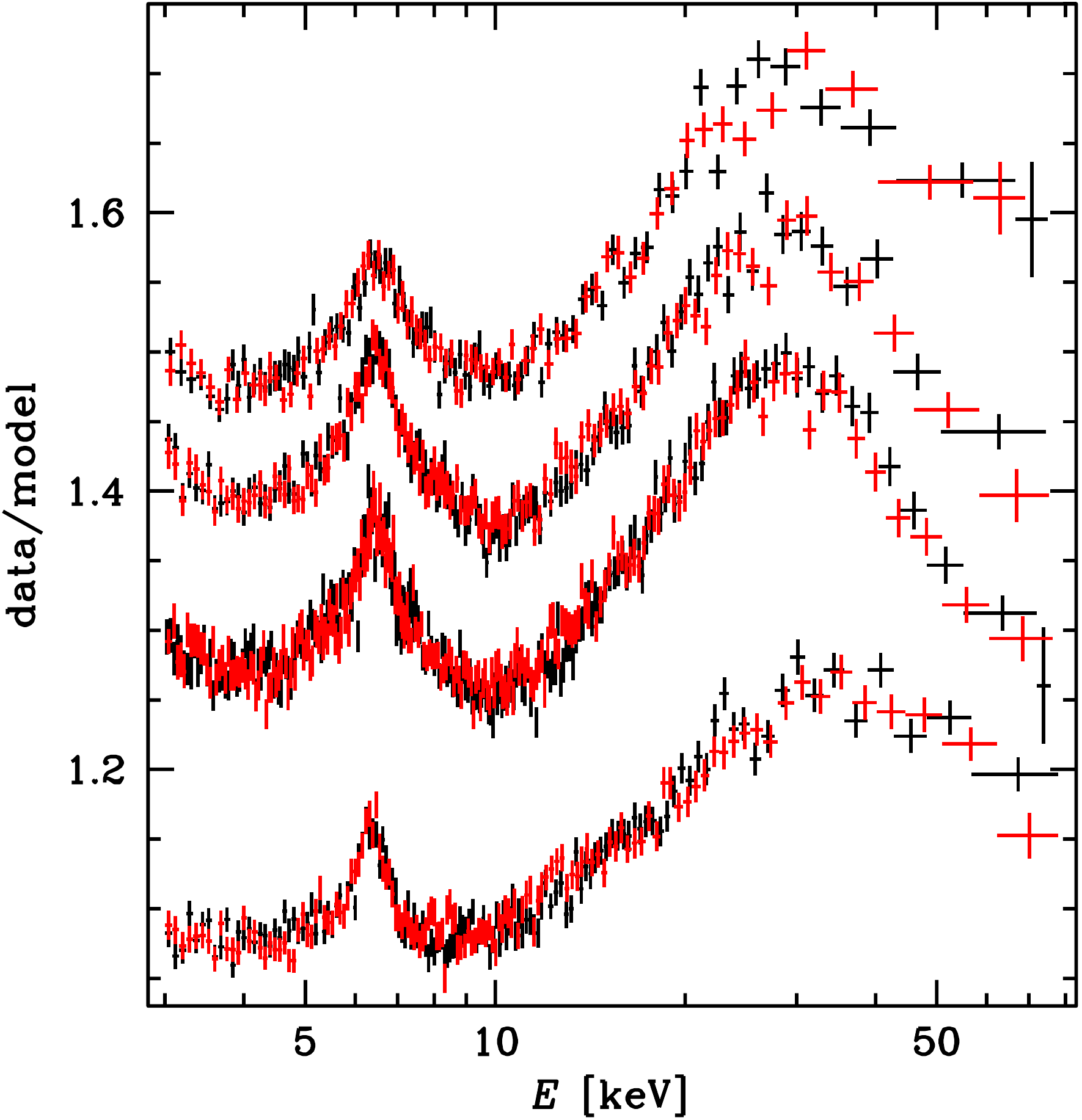}
  \caption{Reflection profiles in the \nustar spectra for epochs 1--4 (from bottom to top) presented as the data/model ratios after removing the two reflection components in the model with separate reflection regions. For clarity of display, the profiles for epochs 3 and 4 have been offset by $+0.15$ and $+0.35$, respectively. We see that the profiles look remarkably similar. Still, the data are fitted with the best-fit values of $R_{\rm in}$ ranging from $\approx 10$ (epoch 4) to $>$100 (epoch 3), see Table \ref{t_fits}.  
}\label{refl_ratio}
\end{figure}

We have found that the hard-state spectra above 3\,keV are well described by a structured accretion flow with the geometry shown in Figure \ref{geo}. The energetically dominant component is the hardest one (except for epoch 3, where the two components have comparable fluxes), but its emission is reflected from remote parts of the disk, $R\gtrsim R_{\rm tr}\sim 10^2 R_{\rm g}$. However, the plasma location at $R\gtrsim R_{\rm tr}$ would disagree with a number of arguments. First, the relative similarity in the bolometric flux between the plateau hard state and the soft state \citep{Shidatsu19} argues against a radiative inefficiency of the hard state. Then, the most luminous component should originate close to the BH, but not at $\gtrsim 10^2 R_{\rm g}$. Second, low-frequency variability of the observed flux at higher energies lags behind that at a lower energies, which phenomenon is called `hard lags'. This has been observed in \source (\kara; \citealt{Wang20_HXMT,DeMarco21}) as well as in other BH XRBs, and it has been interpreted as propagation of fluctuations in the accretion flow \citep{Kotov01}. In this framework, a plasma emitting a harder spectrum should be located downstream that with a softer spectrum. Third, a hard Comptonization spectrum requires a low flux of incident seed soft photons (e.g., \citealt{PVZ18}), implying a plasma location away from the disk, at $R<R_{\rm in}$, as shown in Figure \ref{geo}. Its scale height has to be large, as implied by the typical fractional reflection of ${\cal R}_{\rm h}\sim 0.5$, see Table \ref{t_fits}. Also, the outer disk is likely to be flared, which increases the solid angle subtended by it as seen from the central hot plasma.

Then, the source of the softer X-ray component appears to be at $R> R_{\rm in}$. We propose it forms a corona above the disk. Its softness, with the photon index of $\Gamma\sim 2$, is explained by the re-emission of its flux incident on the disk \citep{HM91,PVZ18}. The underlying disk is strongly ionized by the coronal radiation. The reflection features are mildly relativistic and attenuated by the scattering in the corona. The corona extends out to $R_{\rm tr}$.

Our preferred geometry of the inner accretion flow is similar to that of \citet{Mahmoud19}, shown in their fig.\ 2, inferred from a study of the BH XRB GX 339--4. The main difference in our picture is that we propose the outer hot plasma to form a corona above a disk, while the soft-emitting plasma in \citet{Mahmoud19} is placed downstream of the truncation radius.  Alternatively, the reflection of the soft emission can be from cold clumps within the hot plasma, as in the model for Cyg X-1 of \citet{Mahmoud18b}. In fact, the viscous dissipation within a full disk between $R_{\rm in}$ and $R_{\rm tr}$ is likely to lead, via cooling of the coronal plasma, to spectra softer than those we see in the data. The presence of the underlying cold clumps covering only a fraction of the midplane instead of a full disk would then reduce the cooling, e.g., \citet{PVZ18}. Also, our finding of the spectral complexity, requiring at least two Comptonization components in order to explain the spectra at $>3$\,keV agrees with those of \citet{Chakraborty20} and \citet{Wang20_HXMT} for this source and that of \citet{Zdziarski21} for XTE J1752--223, another transient BH XRB.

Our finding that the reflection features in the hard state are only mildly relativistically smeared (provided the fitted inclinations agree with the observational constraints) can explain the fact, pointed out by \kara for the \nicer data and by \buisson for the \nustar data, that the Fe-K range profiles appear similar over the hard state. This is because the relativistic distortion of the rest-frame Fe K spectra is modest. We show the reflection profiles obtained in our fits in Figure \ref{refl_ratio}, with respect to the sum of the two Comptonization components. While the profile for epoch 1 looks relatively narrow, those for 2, 3 and 4 look similarly broad. Still, we have found that the inner radii for reflection of the hard and soft spectral components do vary a lot across the studied data set. In particular, $R_{\rm in}>100 R_{\rm g}$ for epoch 3 while $R_{\rm in}\approx 10 R_{\rm g}$ for the best fit to the data for epoch 4.  The variability of the characteristic disk radii can readily explain the findings by \kara and \buisson that the reverberation and power-spectrum time scales decrease with the increasing softness in spite of the Fe K profiles looking relatively similar. 

Therefore, the statements of \kara and \buisson that the varying characteristic time scales accompanied by the Fe K complexes looking similar imply strong variations of the coronal scale height above a disk of a constant inner radius (see fig.\ 4 in \kara) appear not certain. The dominant trend found by our spectral fitting is that $R_{\rm tr}$ decreases with the increasing softness of the spectra. Thus, the reverberation time scales will correspond to the distance between the dominant hard-emitting plasma and outer parts of the disk, beyond $R_{\rm tr}$. A contraction/expansion of the corona is still allowed, but it is not required. Given our spectral results, it is unlikely to be the dominant cause of the variable reverberation time scales.

Still, details of the source geometry and its changes with the spectral evolution in the hard state remain unclear. The solid angle subtended by the reflector as seen by the inner hot plasma is large, $\sim\!\! 0.5\times 2\pi$. This requires either a large scale height of that plasma and/or the disk beyond $R_{\rm tr}$ to be flared (see Figure \ref{geo}), but the latter may cause an obscuration of the central hot plasma. A large scale height can be achieved if the accreting flow is outflowing, as in the model of \citet{Beloborodov99, MBP01}. This may form a slow sheath of the jet (e.g., \citealt{Reig21}). Arguments for a part of the X-ray emission of \source to be from the jet are presented by \citet{Wang20_HXMT} and \citet{Ma21}. We also note that \citet{DLHC99} pointed out that while the irradiation of the outer disk is required by the observed light curves of transient BH XRBs, it cannot be achieved in the standard disk model because of self-screening. This requires a geometry in which the solid angle subtended by outer parts of the disk as seen by the central source is large, similar to our finding. Our proposed geometrical model is also incomplete because it does not account for the spectra below 3 keV, see Figure \ref{nustar_nicer_ratio}. Those data indeed imply the presence of additional components in the accretion flow \citep{Dzielak21,DeMarco21}.

\section{Conclusions}

We have confirmed the findings of a number of previous works that the geometry in the hard state of BH XRBs features a significantly truncated disk, and the dominant trend is a decrease of the characteristic disk radii with the decreasing hardness, i.e., during the evolution toward the soft state. The changes of $R_{\rm in}$ occur in spite the visual appearance of a constancy of the Fe K profile (see Figure \ref{refl_ratio}). For $L\sim 5\% L_{\rm E}$ during the rise of \source, the truncation radius was $\sim\! 10^2 R_{\rm g}$, as found with a number of different models. 

Then, we find the accretion flow is structured, with at least two primary Comptonization components, the hard (usually dominant) and soft (see Figure \ref{reflkerr2}). The reflection features are dominated by reflection of the hard components far away in the disk. The soft component is responsible for the broader reflection component, also seen in the data. A possible geometry accounting for that is shown in Figure \ref{geo}. 

Our findings imply that the evolutionary changes of the reverberation and power-spectrum time scales can be explained by changes of $R_{\rm in}$ and $R_{\rm tr}$, without the need for invoking a corona contraction. 

We have also found another family of spectral solutions with the disk extending to an immediate vicinity of the ISCO, confirming \buisson and \citet{Chakraborty20}. However, those solutions require a low inclination, $i\sim 30\degr$, while the inclinations of the binary and the jet have been found to be in the 60--$81\degr$ range. 

The data at $<$3\,keV from \nicer, not modeled in this work, show strong and complex soft X-ray excesses (Figure \ref{nustar_nicer_ratio}), implying the presence of at least one more Comptonization component, in addition to a disk blackbody.

\section*{Acknowledgments}
We thank J. Casares, C. Done, J. Kajava, N. Kylafis, M. Torres, A. Veledina and Y. Wang for valuable comments and discussions, and the referee for valuable comments. Special thanks are due to D. Buisson for his finding the low-radius lamppost solution for the current data of epoch 1. We have benefited from discussions during Team Meetings of the International Space Science Institute (Bern), whose support we acknowledge. We also acknowledge support from the Polish National Science Centre under the grants 2015/18/A/ST9/00746 and 2019/35/B/ST9/03944, and from Ram{\'o}n y Cajal Fellowship RYC2018-025950-I.

\bibliography{allbib}{}
\bibliographystyle{aasjournal}

\label{lastpage}
\end{document}